\documentclass[reprint,eqsecnum,floats,aps,amsmath,amssymb,footinbib,prd,twocolumn, showpacs]{revtex4-1}

\usepackage{latexsym,pifont,textcomp,setspace,xspace}
\usepackage{amsmath,amstext,amssymb,color,graphicx}
\usepackage[utf8]{inputenc}

\newcommand{\derB}{\mathring{B}}
\newcommand{\diff}{\mathrm{d}}
 
\usepackage{color}

\begin{document}

\date{2014/09/09}

\title{Correlations across horizons in quantum cosmology}

\author{Ana Alonso-Serrano}
\email{a.alonso.serrano@iff.csic.es}
\affiliation{Instituto de F\'{\i}sica Fundamental (IFF-CSIC), Serrano 121, 28006 Madrid, Spain}
\author{Luis J. Garay}
\email{luisj.garay@ucm.es}
\affiliation{Departamento de F\'{\i}sica Te\'orica II, Universidad Complutense de Madrid, 28040 Madrid, Spain}
\affiliation{Instituto de Estructura de la Materia (IEM-CSIC), Serrano 121, 28006 Madrid, Spain}

\author{Guillermo A. \surname{Mena Marug\'an}}
\email{mena@iem.cfmac.csic.es}
\affiliation{Instituto de Estructura de la Materia (IEM-CSIC), Serrano 121, 28006 Madrid, Spain}

\begin{abstract}

Different spacetime regions separated by horizons are not related to each other. We know that this statement holds for classical spacetimes. In this paper we carry out a canonical quantization of a Kantowski-Sachs minisuperspace model whose classical solutions exhibit  both an event horizon  and a cosmological horizon in order to check whether the above statement also holds from the quantum gravitational point of view. Our analysis shows that  in fact this is not the case: Quantum gravitational states with support  in spacetime configurations that exclusively describe either the region between horizons or outside them are not consistent in the sense that there exist unitary operators describing a natural notion of evolution that connect them. In other words, unitarity is only preserved in this quantization when dealing with the whole spacetime and not in each region separately.

\end{abstract}

\maketitle

\section{Introduction}

Horizons are ubiquitous in classical General Relativity. Although there are many different types of horizons, the main common consequence of their presence is that they, someway or another, excise parts of spacetime  making them inaccessible from the outside.
This is the case, for instance, for event and cosmological horizons on which we will concentrate in this paper. From our spacetime region, physical observations beyond these horizons are classically out of the question. It has been proposed in the literature~\cite{Jacobson:1991gr,Jacobson:1993hn,Unruh:1994je} that  at very high energies horizons may be blurred because of effective superluminal modifications of the dispersion relations that would allow high energy modes to leak across the classical horizon. These proposals are inspired in the behaviour of analogous configurations in  condensed matter systems. This would solve, for instance, the so-called trans-planckian problem in black hole physics~\cite{Unruh:1976db}. But softening the horizons is a non-trivial task that may affect the global spacetime structure~\cite{Barbado:2011ai}. This could be taken as an indication that horizons might not be as impressive and frightening as they seem from the classical point of view and might allow quantum mechanically for interactions between classically separated regions.

Only upon the fall of these titans, we could try to take one more (huge, granted) step and speculate about other universes than our own. If horizons are keeping us from peeking in regions of our own universe, it will  be even more difficult to try to advance any possible information about the physical events and characteristics of those universes. However, as we will argue in this paper, quantum mechanics applied to the whole spacetime will not only allow but force the opening of those excised regions and connect them  to our own. So this outrageously big leap might not be beyond our capabilities after all. The potential connections with these other universes, being quantum in nature, could be expected to have characteristics similar to those found in the connections across horizons. 
Actually, from the classical point of view and hence subject to the presence of horizons, classical links \cite{Morris:1988tu,Morris:1988cz,Yurov:2006we,GonzalezDiaz:1996sr,GonzalezDiaz:1998pc,Gott:1997pm} (by means of Lorentzian wormholes) and quantum tunnels \cite{Hawking:1988ae,Hawking:1990in,Giddings:1988wv,Garay:1993rr,MenaMarugan:1994qz} (Euclidean wormholes) between otherwise separate universes have been previously considered, also taking into account that they can imprint some observable signatures \cite{GonzalezDiaz:1997xc,Shatskiy:2007xr,Shatskiy:2008ym,GonzalezDiaz:2011ke}. On the other hand, an attempt to pinpoint possible observable quantum effects of this multiverse on a single universe has been considered, in the formalism of third quantization, by means of the possibility of entanglement between pairs of universes \cite{RoblesPerez:2010ut,RoblesPerez:2011yj,RoblesPerez:2011zp,AlonsoSerrano:2012wc}.
Instead of embarking ourselves in this vast task on mostly unexplored territory (quicksand, as Coleman would say \cite{Coleman:1988tj}), we will delve into the quantum theory for a single universe and analyze the role of horizons in it,  a related but simpler endeavor.

With this aim we will consider a minisuperspace model for a spherically symmetric spacetime in the presence of a positive cosmological constant. This minisuperspace model can be written in terms of a Kantowski-Sachs metric that depends on two variables. 
The maximal analytic extension 
of the classical solutions (Schwarzschild-deSitter spacetimes)  typically contain both black hole and cosmological horizons that isolate our spacetime region from those beyond. Quantizations of spacetimes of Kantowski-Sachs type have been analyzed before from different points of view and with different matter contents \cite{Halliwell:1990tu,Campbell:1990uu,Garay:1991um,MenaMarugan:1993rs,MenaMarugan:1994sj} (see also \cite{Kuchar:1994zk,Gambini:2013ooa}).
In this paper we will carry out a canonical quantization procedure that specifically allows us to tackle with the issue of permeability across  the  horizons. We will follow an extension of Dirac’s 
canonical quantization program  for systems (like ours) with first-class constraints \cite{yeshiva} (along the lines developed by Ashtekar et al. \cite{Ashtekar:1994kv}).

We will decompose the physical Hilbert space of the system into two subspaces corresponding to states with support in configurations that exclusively describe spacetime regions either between or beyond the horizons. It will turn out that these Hilbert subspaces are not stable under the action of unitary operators that implement a natural notion of evolution on physical states. The home for this physical evolution will be the tensor product of both subspaces, which hence will not be dynamically separable but entangled. This means that quantum correlations among classically disjoint regions are a generic unavoidable feature in this quantization.  This entanglement between classically disconnected regions opens up the possibility that if we embraced a broader picture in which  our universe is not isolated but multiply connected to others, we might need to consider the complete structure of the multiverse in order to make a quantum theory and that there would be some kind of quantum effects of other universes in our own, driven by entanglement.

The outline of the paper is the following. In Section \ref{sec:class}, we construct the  model and describe the classical solutions to Einstein equations. Once we have the phase space of our system, in Section \ref{sec:canquant} we quantize it following  Dirac's  extended canonical quantization program \cite{Ashtekar:1994kv} and analyze various useful  bases and representations of the physical Hilbert space. Section \ref{sec:quanthor} is devoted to discuss the generic presence of quantum correlations between classically separated regions. We summarize and conclude in Section \ref{sec:concl}.

\section{Classical solutions}
\label{sec:class}

We construct a model with a general spherically symmetric metric
that depends on two variables $A$ and $b$ ---which play the role of
our dynamical  variables  to construct the configuration space--- and on the lapse function $N$:
\begin{align}
\sigma^{-2}\diff s^2=-\frac{N(r)^2}{A(r)}\diff r^2+A(r)\diff T^2 +b(r)^2\diff \Omega_2^2,
\label{eq:metric}
\end{align}
where all metric variables and coordinates are dimensionless, $\diff\Omega_2^2$ is the line element on the unit two-sphere,  and $\sigma:=\sqrt{2G/\int \diff T}$ has units of length, with $G$ being Newton's constant.  
Note that this is nothing but a Kantowski-Sachs metric, with a suitably redefined lapse  \cite{MenaMarugan:1993rs}.

The corresponding curvature scalar (for $N=1$) is
\begin{align}
b^2\sigma^2R=  2+2 A  \dot b^2+b ^2  \ddot A +4 b \dot A\dot  b  +4 bA  \ddot b ,
\end{align}
where the dot denotes derivative with respect to $r$. For a general lapse function, it suffices to replace this derivative with $1/N$ times the dot derivative.

Then, the
Hilbert-Einstein action (up to surface terms) can be written in terms of the metric configuration variables and a cosmological constant $\Lambda$ as
\begin{align}
S&=\frac{1}{16\pi G}\int\diff^4x\sqrt{-g}(R-2\Lambda)
\nonumber\\
&=-\int \diff r \bigg(\frac{A\dot b^2}{N}+\frac{b\dot b\dot A }{N}+N\derB(b) \bigg)+\text{surf. terms},
\end{align}
with $\lambda=\sigma^2\Lambda$ and
\begin{equation}
B(b)=\frac{\lambda}{3} b^3-b,
\qquad
\derB(b) =\partial_bB(b)=\lambda b^2-1.
\end{equation}

Before we continue, let us make a few comments that may be relevant in the rest of this paper.

The action has been written as an integral over the coordinate $r$ on the the patch considered for that coordinate, which is not necessarily all the positive semi-axis.
Also, if we take the square root of the determinant of the metric properly, we see that $N$ should rather be $|N|$ unless we are continuing it analytically. We do so in the following, although had we considered only positive lapses, this subtlety would not have been relevant.

In principle, we take the range of $b$ to be the whole real line. We see that the metric is invariant under a change of sign in $b$. This means that if we did not restrict its value to, say, the positive real axis, every trajectory would be considered twice. We will take this point into account later on. The range of the variable $A$ is also taken to be the whole real line. This choice  is of much importance in our treatment: A change of sign in $A$ corresponds to a change in the character of the radial coordinate from timelike to spacelike or vice versa. Generically, horizons correspond to $A=0$.

It will be convenient for our analysis to use the new variable  \cite{Campbell:1990uu}
\begin{align}
c=Ab
\end{align}
instead of $A$,
which allows us to simplify the action, that now reads
\begin{align}
S= -\int \bigg(  \frac{\dot b\dot c}{N}+N\derB(b) \bigg),
\end{align}
to which the same comments above apply.

The variational principle for this action gives the classical equations of motion
\begin{align}
(\dot c/N)^{\cdot}&=2\lambda Nb, \qquad \dot b\dot c=N^2\derB(b) ,\nonumber \\
(\dot b/N)^{\cdot}&=0.
\end{align}
The general solution to these equations is
\begin{align}
\dot b=\alpha N, \qquad
\alpha^2 c= B(b) +2m,
\label{eq:classsolbc}
\end{align}
$\alpha$ and $m$ being integration constants. From the point of view of the metric (\ref{eq:metric}), $\alpha$ amounts to a constant rescaling of the coordinates $r$ and $t$. This solution corresponds to the Schwarzschild-(anti)de Sitter metric: Indeed, for $\alpha=N=1$, we have
\begin{align}
b(r)=r,\qquad A(r)=-1+\frac{2m}{r}+\frac{\lambda r^2}{3},
\end{align}
and the horizons are located at the zeros of $A(r)$.

From now on we will only consider the case with positive cosmological constant $\lambda>0$. Negative cosmological constant scenarios can also be treated in an entirely analogous manner. 
 
\begin{figure}
\includegraphics[width=.32\columnwidth]{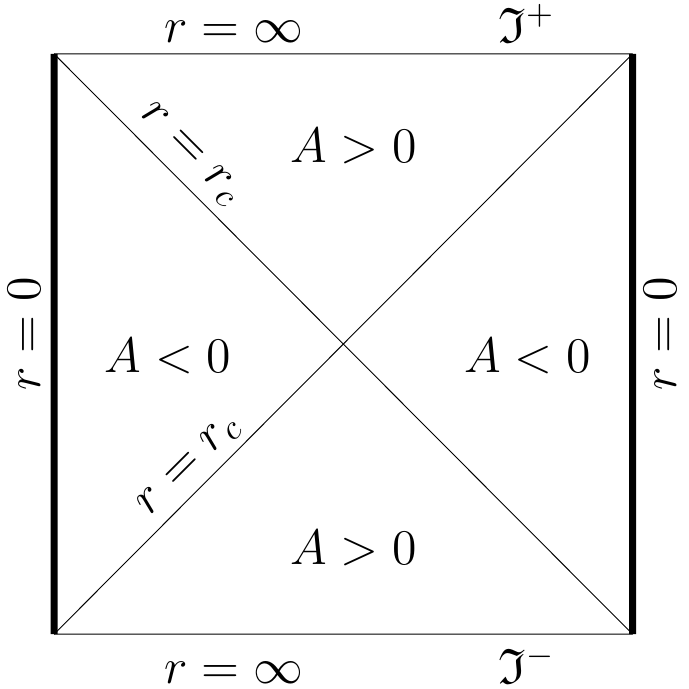}
\includegraphics[width=.32\columnwidth]{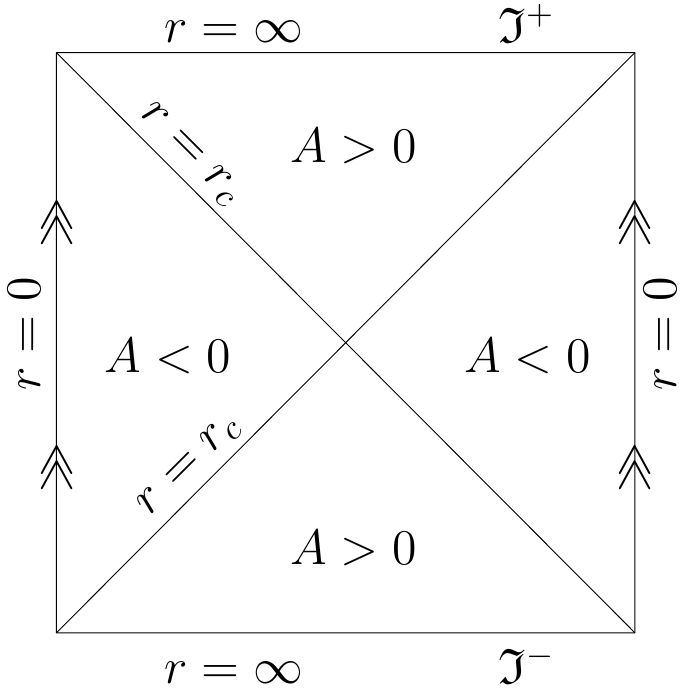}
\includegraphics[width=.32\columnwidth]{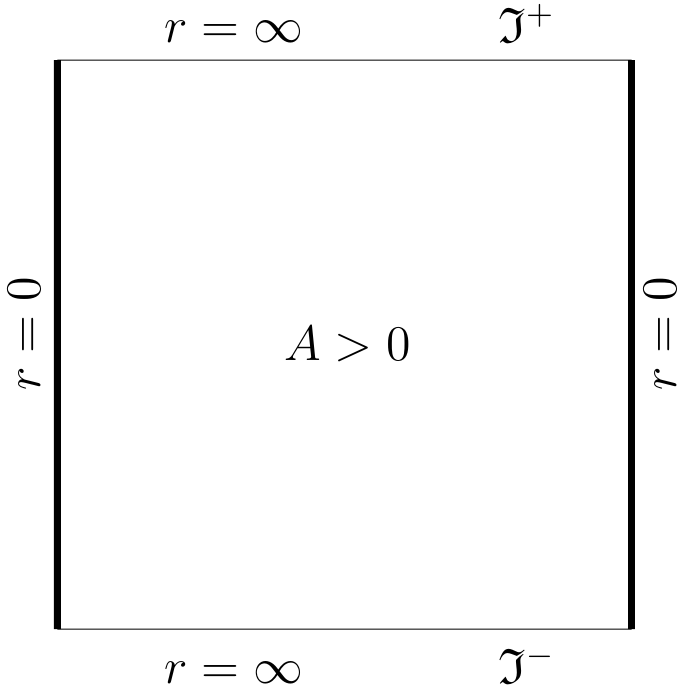}\\
\includegraphics[width=.45\columnwidth]{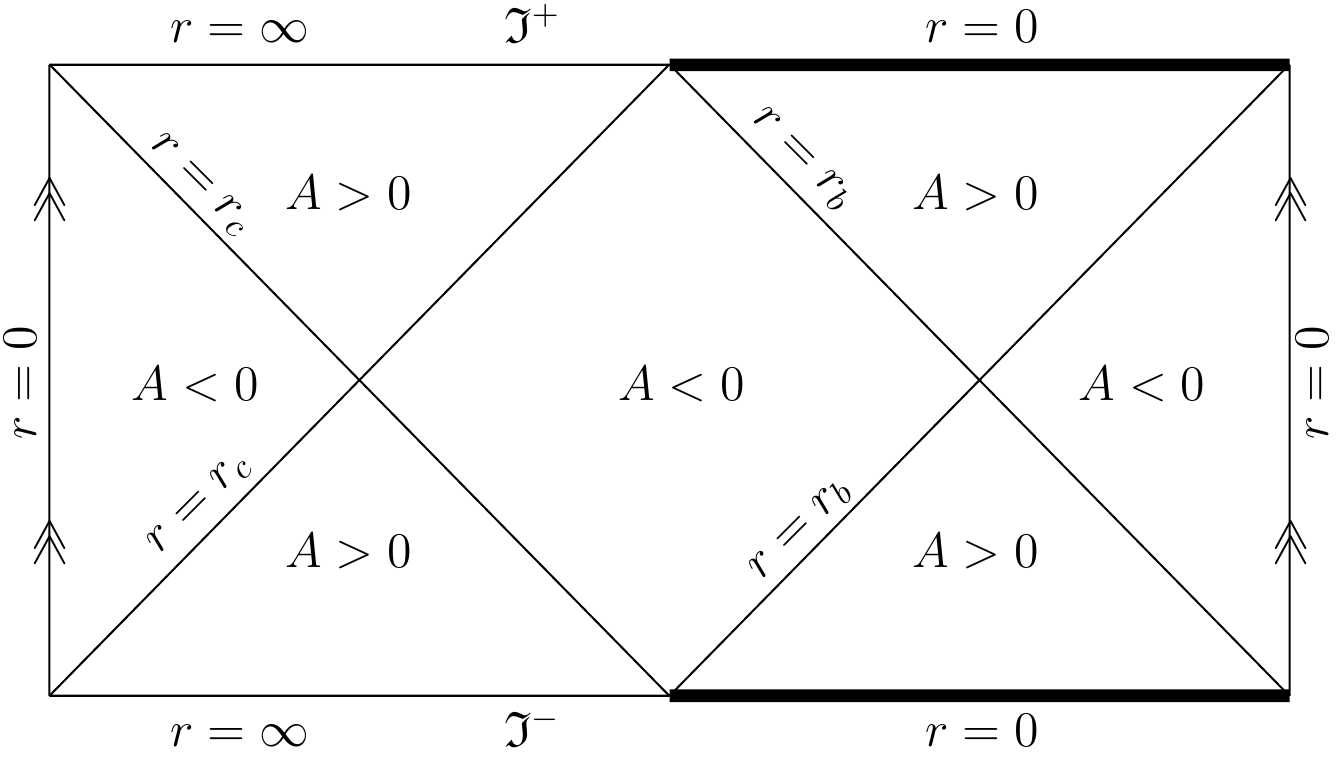}
\includegraphics[width=.53\columnwidth]{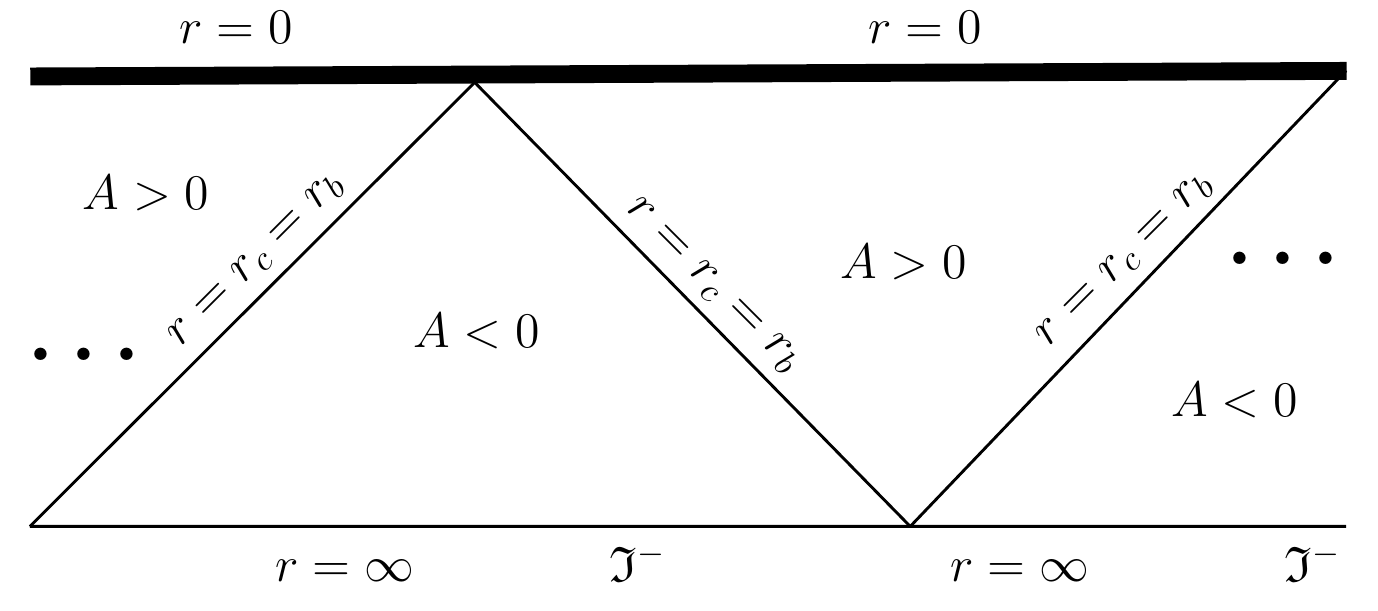}

\caption{Penrose diagrams for the different classical solutions (from left to right and from top to bottom, $m<0$, $m=0$, \mbox{$1/\sqrt{9\lambda}<m$}, \mbox{$0<  m <1/\sqrt{9\lambda}$}, and    \mbox{$m=1/\sqrt{9\lambda}$}). Double arrows indicate identification of the corresponding lines and thick lines represent singularities. Depending on the  mass there are one or two horizons (or none, with a  naked singularity). In all cases, the region for an observer like us is characterized by $A<0$.}
\label{fig:penrose}
\end{figure}

The causal structure of  these spacetimes  is well known~\cite{Lake:2005bf}. There exist different cases depending on the value of $m$. All of them (except for $m=0$) present a singularity at $r=0$. We can see the diagrams for the different cases represented in Fig. \ref{fig:penrose}. The most interesting case is  $0<  m <1/\sqrt{9\lambda}$. We recall that, then, $A(r)=0$ has two positive solutions, at which there are  two horizons:  a black hole horizon (denoted by $r_b$) and a cosmological horizon (denoted by $r_c$). 

A common feature to all solutions that allows us to characterize ``our'' spacetime region (between horizons) in contrast with the regions outside them is the sign of the variable $A$: It is negative inside and positive outside. 

Note that although the metric configuration variables $b$ and $c$ belong to ${\mathbb{R}}$, the range $b\in {\mathbb{R}}_+$ is preserved by the dynamics and so is $\alpha^2c-B\in {\mathbb{R}}_+$. In other words, these ranges are not related to other (negative) values outside them by classical solutions. Then, classically we have the different regions totally disconnected from each other. For the time being we will keep both ranges to be the whole real line.
  
In order to perform a canonical quantization, we are interested in making a Hamiltonian formulation of the system. The canonical action can be expressed as
\begin{equation}
S= \int dr \left(
\dot{c}p_c+\dot{b}p_b-NC\right),
\end{equation}
where the canonical conjugate momenta are
\begin{align}
p_b=-\frac{\dot c}{N},\qquad
p_c=-\frac{\dot b}{N},
\end{align}
and the variation with respect to the lapse function gives rise to the Hamiltonian constraint $C=0$, with
\begin{align}
C= -p_bp_c+\derB(b).
\end{align}

Note that $p_c$ commutes with $C$ under Poisson brackets and therefore is a constant of motion. It is also easy to see that the Hamiltonian constraint of the system and the (classical) metric are invariant under simultaneous changes of sign in the momenta. This can be interpreted as a reversal in the evolution, corresponding to a change of sign in the lapse function $N$. We can remove this reversal considering only positive $N$. The system also has the symmetry  $(b,c,p_b,p_c) \rightarrow (-b,-c, p_b,  p_c)$.
This symmetry could be used to reduce the relevant part of phase space to half of it (in this sense, the duplicity of trajectories with a different sign of $b$ would be removed). We will not impose it; instead we will use a related symmetry that we will discuss in   section \ref{sec:kinspace} with similar result in reducing to a half the relevant part of the phase space.

\section{Canonical quantization}
\label{sec:canquant}

In order to quantize our simple system, we will first construct a kinematical operator algebra starting from its phase space, and closed under Poisson brackets [36]. Then we will represent this algebra by operators acting on a kinematical complex vector space, which for convenience will be endowed with a Hilbert space structure. 
The Hamiltonian constraint will be represented as a specific operator acting on the kinematical space. The space of physical states will be supplied by the kernel of this constraint and the physical operators will be obtained as elements of the kinematical algebra which map the physical space to itself. Finally, the inner product in this physical space will be determined by requiring that a complete set of real classical observables be represented by self-adjoint operators. 

\subsection{Kinematical space and operator algebra}
\label{sec:kinspace}

We will start with the kinematical algebra constructed from the canonical variables $b,c,p_b$ and $p_c$ (and the unit constant), which is obviously closed under Poisson brackets. As kinematical space we choose the vector space spanned by simultaneous solutions to the equations
\begin{align}
-i\partial_c\Psi_{hp}=p\Psi_{hp},
\qquad
[\partial_c\partial_b+\derB(b)]\Psi_{hp}=h\Psi_{hp},
\end{align}
with $h$ and $p$ being real. These solutions (labeled by the parameters $h$ and $p$) depend on the variables $b$ and $c$ and
 have the form  
\begin{align}
\Psi_{hp}(b,c)= e^{ipc+i[B(b)-bh]/p},
\end{align}
where the singularity at $p=0$ should not be of relevance in the Hilbert space constructed below, since it will have zero measure.

More explicitly any kinematical state will be a linear combination of these solutions, i.e.
\begin{align}
\Psi(b,c)=\int_{\mathbb{R}} \diff h\int_{\mathbb{R}}\diff p \tilde\Psi(h,p)\Psi_{hp}(b,c),
\label{eq:kinstates}
 \end{align}
where $\tilde\Psi(h,p)$ is a distribution. This construction endows the kinematical space  with a complex vector space structure. Actually we have constructed two kinematical representations that we can use: The metric $(b,c)$-representation and the $(h,p)$-representation. 
 
In the metric representation, we represent the kinematical algebra by operators acting as 
\begin{align}
\hat b=b ,
\qquad 
\hat c=c,
\qquad
\hat p_b=-i\partial_b,
\qquad
\hat p_c=-i\partial_c.
\label{eq:standardmetricoprep}
\end{align}
Then, in the $(h,p)$-representation, these operators act as
\begin{align}
\hat b&=-ip\partial_h,
\nonumber\\
\hat c  &= i\partial_p+\frac{B(-ip\partial_h)}{p^2}+\frac{i\partial_h h}{ p},
\nonumber\\
\hat p_b&=\frac{\derB(-ip\partial_h)-h}{p},
\nonumber\\
\hat p_c&=p ,
\label{eq:metricops1}
\end{align}
as can be checked by direct application of these operators on the kinematical states (\ref{eq:kinstates}). We have assumed that integration by parts can be carried out without boundary contributions, thanks to the boundary conditions implied by the fact that the states belong to the kinematical Hilbert space (determined later on). Also, a  factor order has been chosen in the last term of the operator  $\hat c$, so that $h$ acts on the right of $\partial_h$ (the reason for choosing this factor ordering will soon be apparent).

An alternative equivalent way of constructing the same  kinematical space can be followed by choosing the canonical set of variables ($t$, $q$, and their corresponding momenta $h,p$) adapted to the system studied in Ref. \cite{MenaMarugan:1993rs}, given by
\begin{align}
t&=-\frac{b}{p_c} ,
\nonumber\\
h&= - p_bp_c+\derB  ,
\nonumber\\
q&=c-\frac{B(b)+ bp_bp_c-b\derB(b)}{p_c ^2},
\nonumber\\
p&=p_c.
\label{eq:cantrans}
\end{align}
The type-2 generating function for this invertible one-to-one canonical transformation on phase space is
\begin{align}
F(c,b,h,p)=cp+\frac{B(b)-bh}{p}.
\end{align}
From the classical point of view,  $p$, $q$, and $h$ are constants of motion. In fact, by comparison with the classical solution (\ref{eq:classsolbc}) in terms of the metric variables $b$ and $c$ we see that
\begin{align}
p=\alpha,
\qquad
h=0,
\qquad
q=\frac{2m}{\alpha^2},
\qquad \dot t=-N.
\end{align} 
Notice that $q$ is positive on classical solutions if we want to consider only positive mass (i.e. absence of naked singularities). 
It is also interesting to note that $q$ is just the value of the the dynamical variable $c$ at $b=0$ (the relevance of this comment will become apparent when analyzing quantum representations in the next section). The  symmetry discussed at the end of the previous section,
\begin{equation}
(b,c,p_b,p_c) \rightarrow (-b,-c, p_b,  p_c),
\end{equation}
in terms of these new canonical variables, now becomes  the symmetry  \begin{align}
(q,t,p,h) \rightarrow (-q,-t, p,  h).
\end{align} 
This symmetry implies that all the relevant information is actually contained in half of the original phase space and that, in consequence,  we can restrict
our study to it. Since we are interested for other reasons on positive $q$, this is the half that we will choose. We will discuss how to impose this symmetry as a restriction on the wave functions later on (the corresponding representation will be given by the restriction of the ``complete'' representation to a subspace). For the time being, we will keep it unrestricted.

We choose as kinematical vector space the space of distributions $\tilde\Psi(h,p)$ and represent the kinematical algebra on it as
\begin{align}
\hat h=h,
\qquad
\hat t=i\partial_h,
\qquad
\hat p=p,
\qquad
\hat q=i\partial_p.
\end{align} 
The metric variables can be represented as the operators
\begin{align}
\hat b&=-\hat t\hat p,
\nonumber
\end{align}
\begin{align}  
\hat c&=\hat q+B(-\hat t\hat p){\hat p}^{-2}+\widehat{th}{\hat p}^{-1},
\nonumber\\
\hat p_b&=[\derB(-\hat t\hat p)-\hat h]{\hat p}^{-1},
\nonumber\\ 
\hat p_c&=\hat p,
\label{eq:metricops}
\end{align}
as can be easily seen by inverting the canonical transformation (\ref{eq:cantrans}).
For convenience, in contrast with Ref. \cite{MenaMarugan:1993rs} and in agreement with the operator order chosen in Eq. (\ref{eq:metricops1}), we take  $\widehat{th}=\hat t\hat h$, even if it is not a symmetric ordering,  so that the action of this operator on physical states (which are annihilated by $\hat h$, as we will see) vanishes. 

To make contact with the construction presented in the beginning of this section, we can go to the metric $(b,c)$-representation  by means of the transformation
\begin{align}
\Psi(b,c)=\int_{\mathbb{R}} \diff h\int_{\mathbb{R}}\diff p \tilde\Psi(h,p)e^{iF(b,c,h,p)}.
\label{eq:transform}\end{align}
We can see that this expression is precisely that in Eq. (\ref{eq:kinstates}).
In this metric representation, the metric canonical variables are represented as the operators given in Eq. (\ref{eq:standardmetricoprep}). It is also worth emphasizing that we are  using a slightly different representation in comparison with that in Ref. \cite{MenaMarugan:1993rs}.

It may be convenient to introduce an inner product in the kinematical space on which the operators $\hat h$, $\hat p$, $\hat t$, and $\hat q$ are self-adjoint, namely:
\begin{align}
(\Psi_1,\Psi_2)=\int_{\mathbb{R}}\diff h\int_{\mathbb{R}}\diff p\tilde\Psi_1(h,p)^*\tilde\Psi_2(h,p),
\label{eq:innerprod}
\end{align}
where the symbol $*$ denotes complex conjugation.
Then the kinematical Hilbert space is the completion in this inner product of the space of distributions $\tilde \Psi(h,p)$, that is to say $L^2(\mathbb{R}^2,\diff h\diff p)$.
The states $\Psi_{hp}(b,c)$
are obviously orthonormal in the Dirac-delta sense:
\begin{align}
(\Psi_{hp},\Psi_{h'p'})=\delta(h-h')\delta(p-p').
\end{align}

The restriction to positive $q$ can be taken by going to the Fourier transform of the $(h,p)$ representation in the $p$ variable (to $q$) and projecting to the positive semi-axis of the configuration space [the corresponding space of wave functions are those in $L^2(\mathbb{R}\times \mathbb{R}^+,\diff h \diff q)$, obtained by restricting those functions to positive $q$ and using the inner product induced from Eq.
(\ref{eq:innerprod}); this space is not stable under the operator $\hat{p}$ but it is stable under the operator $\widehat {qp}=i(p\partial_p+1/2)$ instead (that is self-adjoint)]. Nonetheless, we will continue to consider the general Hilbert space without restricting $q$, keeping in mind that this implies a physical duplicity, as discussed above.

Finally, the Hamiltonian constraint can be represented in this kinematical space by the operators
\begin{align}
\hat C=\partial_b\partial_c+\derB(b),
\qquad
\hat C=h,
\end{align}
in the metric $(b,c)$-representation and in the $(h,p)$-representation, respectively.

\subsection{Physical Hilbert space}

In the $(h,p)$-representation on the kinematical space,  the Hamiltonian constraint is represented as multiplication by $h$, as we have just seen. Therefore, the space of solutions can be obtained by solving the equation
\begin{align}
\hat C\tilde \Phi(h,p)=h\tilde \Phi(h,p)=0.
\end{align}
The solutions have the form
\begin{align}
\tilde\Phi(h,p)=\frac{1}{\sqrt{2\pi}}\delta(h) \phi(p),
\end{align}
where $\phi(p)$ is an arbitrary  distribution and the constant prefactor has been chosen for normalization purposes.
In this physical vector space, the operators $\hat p$ and $\hat q$ (which commute with the constraint $\hat C$)  are represented as 
\begin{align}
\hat p=p,\qquad
\hat q=i\partial_p.
\end{align}

The remaining task in the canonical quantization procedure is fixing the inner product  in the space of physical states. We choose it so that the observables $\hat p$ and $\hat q$ be self-adjoint, which leads to the inner product
\begin{align}
\langle\Phi_1,\Phi_2\rangle=\int_{\mathbb{R}}\diff p\phi_1(p)^*\phi_2(p).
\label{eq:physinner}
\end{align}

To summarize, the physical Hilbert space of quantum states for our system is $L^2(\mathbb{R},\diff p)$, which contains just one degree of freedom as expected. We will refer to this representation as the $p$-representation.  

The  $p$-representation is not the only representation that we can use and in fact there exist other representations of physical interest as we will see.
Let us note that physical states can also be written in terms of   the metric variables $b$ and $c$ as 
\begin{align}
\Phi(b,c)&=\int_{\mathbb{R}} \diff h\int_{\mathbb{R}}\diff p \tilde\Phi(h,p)e^{iF(b,c,h,p)}
\nonumber\\
&=\frac{1}{\sqrt{2\pi}}\int_{\mathbb{R}} \diff p \phi(p)e^{i[pc+B(b)/p]},
\end{align}
the inverse of this transformation being 
\begin{align}
\phi(p)=\frac{1}{\sqrt{2\pi}}e^{-iB(b)/p}\int_{\mathbb{R}}\diff c\Phi(b,c)e^{-ipc}.
\end{align}
Note that the dependence of physical states $\Phi(b,c)$ on $b$ is only through $B(b)$. 

The closure relation in terms of the metric variables can then be easily obtained:
\begin{align}
\openone(b,c;b,c')&=\frac{1}{2\pi}\int_{\mathbb{R}}\diff p e^{ip(c-c')}e^{i[B(b)-B(b)]/p}
\nonumber\\ &
=\delta(c-c'),
\end{align}
so that
\begin{align}
\Phi(b,c)=\int_{\mathbb{R}}\diff c'\openone(b,c;b,c')\Phi(b,c').
\end{align}

Finally, we can write  the inner product in terms of the metric variables:
\begin{align}
&\langle\Phi_1,\Phi_2\rangle=\int_{\mathbb{R}}\diff p\phi_1(p)^*\phi_2(p)
\nonumber\\
&=\frac{1}{2\pi}\int_{\mathbb{R}}\diff p \int_{\mathbb{R}}\diff c_1\int_{\mathbb{R}}\diff c_2 e^{ip(c_1-c_2)}\Phi_1(b,c_1)^*\Phi_2(b,c_2)
\nonumber\\
&= \int_{\mathbb{R}}\diff c \Phi_1(b,c)^*\Phi_2(b,c).
\label{eq:innerbc}
\end{align}

After these preliminary notes about changes of representation, let us analyze  some equivalent representations  that will be particularly adequate to the study of horizon quantum physics that we want to carry out. 

\subsection{Equivalent representations}
 
From the above discussion, it immediately follows that we can change from the $p$-representation to the metric  representation  for physical states $\Phi(b,c)$ by  means of a Fourier transform  (together with a multiplication by a $b$-dependent phase).
 This fact allows us to introduce two other families of representations.

\subsubsection{$c_b$-representations}

We have seen that we actually have not only one but a whole family of $c_b$-representations labeled by $b$. This is obvious in the   formula (\ref{eq:innerbc}) for the inner product, valid for any value of $b$. This resembles a kind of transformation from the Heisenberg picture, in which the states $\phi(p)$ only depend on the $p$, to the $b$-Schrödinger picture, where the states now depend on $p$ and $b$ (the $b$-evolution being driven by the Hamiltonian $-\derB(b)\hat p^{-1}$), together with a Fourier transform to the variable $c$. In this sense, we can write 
\begin{align}
\Phi(b,c)=\hat U(b)\Phi(\tilde b,c),
\end{align}
where $\tilde b$ represents any of the roots of the polynomial $B(b)$ and 
\begin{align}
\hat U(b)=e^{iB(b)\hat p^{-1}}.
\end{align}

In each of these $c_b$-representations, the observables that we want to represent will be the  ones corresponding to $\hat p $ and $\hat q$ in this $b$-Schrödinger picture, 
\begin{align}
\hat \pi_b&=\hat U(b)\hat p \hat U^\dag(b)=\hat p,
\nonumber\\
\hat c_b&=\hat U(b)\hat q \hat U^\dag(b)=\hat q+B(b)\hat p^{-2}.
\end{align} 
It is straightforward to see that the action of these canonically conjugate observables on $\Phi(b,c)$ is just derivation and multiplication by $c$, respectively, i.e.
\begin{align}
\hat \pi_b=-i\partial_c,
\qquad
\hat c_b=c,
\end{align} 
and hence the name $\hat c_b$ instead of $\hat q_b$ (note that we have already mentioned this point when we defined the canonical variable $q$). So, we have a family of observables $\hat c_b $, each in a different $c_b$-representation (labeled by $b$), that can be interpreted as giving the value of the metric variable $c$ at the considered value of $b$. 
This interpretation is actually based on the observation made above that $\hat c_b$ is nothing but a kind of Schrödinger picture operator obtained from $\hat q$ by means of the ``$b$-evolution'' operator $\hat U(b)$.

\subsubsection{$p_b$-representations}

In the same way, we also have   a family of $p_b$-representations labeled by $b$, given by the Fourier transform in $c$ of $\Phi(b,c)$:
\begin{align}
\phi(b,p)&=\frac{1}{\sqrt{2\pi}}\int_{\mathbb{R}}dc e^{-ipc}\Phi(b,c)
\nonumber\\
&=\hat U(b)\phi(p)=\phi(p)e^{iB(b)/p},
\end{align}
for which the inner product (\ref{eq:physinner}) reads
\begin{align}
\langle\Phi_1,\Phi_2\rangle=\int_{\mathbb{R}}\diff p\phi_1(b,p)^*\phi_2(b,p).
\label{eq:innerbp}
\end{align}
In these $p_b$-representations, for each $b$, the  operator \mbox{$\hat \pi_b=\hat p$}  acts by multiplication and it is clearly an observable, well defined on (a dense domain of) the physical Hilbert space.  On the other hand, the operator $\hat c_b$  is also an observable as we have seen and acts as
\begin{align}
\hat c_b=i\partial_p.
\end{align}

\subsubsection{Schrödinger and Heisenberg pictures}

We can adopt two alternative viewpoints analogous to the Schrödinger an Heisenberg pictures of standard quantum mechanics. Although we will refer to the $p_b$-representations, an entirely analogous discussion holds for the $c_b$-representations.

From the  first  point of view, we can consider that the $p_b$-representations [with states described by $\phi(b,p)$] give the evolution of the $p$-representation [with states described by $\phi(p)$] from a value $\tilde b$ of $b$ where $B(\tilde b)$ vanishes to the new value of $b$ [and therefore of $B(b)$], providing a whole family of representations that give the corresponding Schrödinger ``dynamics'' in the parameter $b$. Notice that, for this, it is not necessary that $B(b)$ be monotonous in $b$. The Hamiltonian of the evolution in $b$, from this viewpoint, would be $-\derB(b)\hat p^{-1}$ so that, when the associated Schrödinger equation is integrated, one gets the phase $iB(b)/p$, as we have discussed. Note that this Hamiltonian is $b$-dependent, and moreover not strictly positive; hence, indeed, the phase $iB(b)/p$ is not monotonous in $b$. Nonetheless, the evolution is unitary, since the inner product~(\ref{eq:innerbp}) is conserved, i.e. $b$-independent.

From the second point of view, we can choose a fixed $p_b$-representation for a given value of $b$ [with states described by $\phi(b,p)$], and represent our family of observables in a kind of Heisenberg picture (see e.g. \cite{Ashtekar:2006uz} for similar definitions of observables). The family of observables corresponding to $c$ at different values $b_0$ of $b$ would be given, in this way, by
\begin{align} 
\hat c_b^{0}=\hat U^\dag(b,b_0)\hat c_b\hat U(b,b_0)=
i\partial_p +\frac{B(b_0)-B(b)}{p^2},
\end{align}
where $\hat U(b,b_0)=e^{i[B(b)-B(b_0)]\hat p^{-1}}$.
This observable gives in the $p_b$-representation the value of $c$ when $b=b_0$. Since the function $B$ is not one-to-one, the operators in this family may coincide for some values of $b_0$, namely those where $B(b_0)$ is the same.  

\subsection{Some bases of the physical Hilbert space}

Before proceeding to our main discussion, which faces the question which motivated our analysis, let us complete our study of the quantization with the determination of some especially useful bases for the physical Hilbert space of our system.
 
 Let us start by considering the following states
\begin{align}
\phi_p(p')=\delta(p-p')
\end{align}
in the $p$-representation. Their counterparts in the $p_b$-representations  are straightforward to find:
\begin{align}
\phi_p(b,p')=\delta(p-p')e^{iB(b)/p}.
\end{align}
They are obviously eigenstates of $\hat p$ with eigenvalue $p$ and, hence, they provide an orthonormal basis. Their counterparts in the $c_b$-representations   are
\begin{align}
\Phi_p(b,c)=\frac{1}{\sqrt{2\pi}}e^{ipc+iB(b)/p},
\end{align}
and the closure relation in these representations reads
\begin{align}
\openone(b,c;b,c')=\int_\mathbb{R} \diff p  \Phi_p(b,c)\Phi_p(b,c')^*=\delta(c-c').
\end{align}

Finally, there is still another family of bases that will prove very helpful in our analysis, namely that made of eigenstates of the self-adjoint operator $\hat c_b^{ 0}$ in the $p_b$-representation with real eigenvalues $c^{ 0}$:
\begin{equation}
\phi_{c^{ 0}}(b,p)=\frac{1}{\sqrt{2\pi}}e^{-ipc^{ 0}-i[B(b_0)-B(b)]/p}.
\end{equation}

In this family of representations, the identity operator acquires the form $\openone(b,p,b,p')=\delta(p-p')$ and can be decomposed as a sum over all eigenvalues $c_0$ (the whole real line) of $\hat c_b^0$ in the following manner
\begin{align}
\openone(b,p;b,p')=\int_{\mathbb{R}}\diff c^{ 0}\phi_{c^{ 0}}(b,p)\phi_{c^{ 0}}(b,p') ^* .
\end{align}
We can decompose this identity in the sum of two orthogonal projectors: one for positive eigenvalues of $c^0$, $\hat P_+^{0}$, and the other for negative eigenvalues, $\hat P_-^{0}$ (the integral over the real line is the sum of the two corresponding half-infinite intervals): 
\begin{align}
\openone=\hat P_+^{0} +\hat  P_-^{0}.
\end{align} 
The superindex ${0}$ makes manifest the dependence of the projection operators on the value of $b_0$  where $c$ is evaluated.
Explicitly, these projection operators can be written as
\begin{align}
\hat P_{\pm}^{0}\phi(b,p) &=\frac{1}{2\pi}\int_{\mathbb{R}^{\pm}}dc^0 e^{-ipc^0-i[B(b_0)-B(b)]/p}\nonumber\\&\times\int_{\mathbb{R}}dp^{\prime} e^{ip^{\prime}c^0+i[B(b_0)-B(b)]/p^{\prime}} \phi(b,p^{\prime}).
\end{align}

\section{Quantization and horizons}
\label{sec:quanthor}

Assume that, at a certain positive value $b_0$ of $b$, we observe only the region with negative values of $c$. This corresponds classically to considering only our region of the universe, i.e. the spacetime region that lies between the black-hole and the cosmological horizons at the given ``instant of dynamical variable'' $b_0$. Similarly, we could restrict ourselves to the exterior of our region of the universe (beyond the black hole and cosmological horizons), i.e. to positive values of $c$ at $b_0$. In our scheme these restrictions can be accomplished by choosing states with null projection under $\hat P_{\pm}^0$, respectively, or equivalently by projecting an arbitrary state with $\hat P_{\mp}^0$ and normalizing the result.

Classically whatever happens beyond the horizons will have no effect whatsoever in our spacetime region. We are now ready to ask ourselves, and also answer, the corresponding quantum mechanical question. More explicitly, the question that we want to address now is whether this restriction to our region of the universe (i.e. to negative values of $c$) is robust and meaningful, so that we can sensibly forget about the regions beyond the horizons quantum mechanically.

If this were not the case, then  measurements of $c$ at a different positive value $b_1$ of $b$ would lead to contradictory results. The question is then whether observations of the values of $c$ at different values  of $b$ are compatible. 
If they were not, the two projections (at the different values $b_0$ and $b_1$) would differ, the corresponding observables $\hat c_b^0$ and $\hat c_b^1$  could not be diagonalized simultaneously and, hence, the eigenstates could not be chosen as common to both observables. In this case, as mentioned above, the projectors would not commute and the restriction to our region of the universe between horizons would not be stable, in the sense that the projection at $b_0$ on  negative values of $c^0$ would generally have a non-vanishing projection at $b_1$ on positive values of $c^1$, and vice versa. The restriction to the interior of the horizons would depend on the value of $b_0$, and would therefore be unstable under evolution in this variable.

To summarize, quantum stability and robustness of the restriction to our region of spacetime requires that  the two considered observables $\hat c_b^0$ and $\hat c_b^1$ commute. We are going to prove that this is not the case, i.e. that $\hat c_b^0$ and $\hat c_b^1$ are not commuting observables. 
A direct calculation shows that
\begin{align}
[\hat c_b^0,\hat c_b^1]&=\bigg[\hat q +[B(b_0)-B]{\hat p}^{-2} ,\hat q +[B(b_1)-B]{\hat p}^{-2} \bigg]\nonumber\\&
=-2i[B(b_1)-B(b_0)]{\hat p^{-3}}\neq 0,
\end{align}
and therefore the family of considered observables are not mutually compatible. Alternatively, this same result can be obtained  if we act with $\hat c_b^1$ on the eigenstates of $\hat c_b^0$. In the $p_b$-representation, it is straightforward to get:
\begin{equation}
\hat c_b^1\phi_{c^0}(b,p)=\bigg[ c^0 - \frac{B(b_0)-B(b_1)}{p^2}\bigg]\phi_{c^0}(b,p).
\end{equation}
We then see that the sector of positive values of $c^0$ (at positive $b_0$) would be contained in the positive sector of $\hat c_b^1$ (at positive $b_1$) if $B(b_0)-B(b_1)<0$, and the sector of negative values of $c^0$ in the negative sector of $\hat c_b^1$ if $B(b_0)-B(b_1)>0$. Both conditions are incompatible unless $B(b_0)=B(b_1)$, which is not satisfied for general values $b_0$ and $b_1$ (maybe just at some points $b$, but not in full intervals). This further supports the conclusion that the projectors at positive and negative $c^0$ at different values $b_0$  of $b$ are not mutually compatible in general.

Note that the operator $\hat q$  can be considered as a particular case of ${\hat c}_b^0$, namely the one associated with $b_0=\tilde b$ [with $B(\tilde b)=0$]. Even if we restrict to positive $q$ by acting with the associated projection to the positive part of the spectrum of this operator (removing in this way the physical duplicity that we were maintaining till now), the system will develop contributions to the negative sector of $\hat q$ for other values of the variable $b$, in accordance with our discussion above. 

Since the dynamics in $b$ mixes the projections, as we have seen, 
describing states as direct sums of positive and negative $c^0$-states is not the best strategy. Instead, it is more appropriate to consider general physical states belonging to the tensor product 
\begin{align}
{\cal H}^0={\cal H}_+^0\otimes{\cal H}_-^0
\end{align} 
of the projection subspaces
\begin{align}
{\cal H}_\pm^0=\hat P_\pm^0 {\cal H},
\end{align}
where, as before, the superindex 0 denotes the choice of a particular instant of $b$ for the construction, and ${\cal H}$ is the Hilbert space from which we started. Using that the sum of $P_+^0$ and $P_-^0$ is the identity, any observable $\hat O$ can then be decomposed in four operators between both projection subspaces:
\begin{align}
\hat O_{\pm\pm}^0:{\cal H}_\pm^0\to{\cal H}_\pm^0,
\qquad
\hat O_{\pm\mp}^0:{\cal H}_\pm^0\to{\cal H}_\mp^0,
\end{align}
defined as
\begin{align}
\hat O_{\pm\pm}^0=\hat P_\pm^0\hat O \hat P_\pm^0,
\qquad
\hat O_{\pm\mp}^0=\hat P_\pm^0\hat O \hat P_\mp^0.
\end{align}

The operators $\hat O_{\pm\mp}^0$ mix the two subspaces ${\cal H}_\pm^0$ corresponding to the considered projections and cause correlations between them. This is the case of $\hat c_b^1$ [for $B(b_1)\neq B(b_0)$)], as we have seen. Moreover, if $\hat O$ is a unitary observable, the existence of the two mixing components will indicate that unitarity is not respected in each of the subspaces ${\cal H}_\pm^0$ separately. Our system certainly exhibits this kind of unitary operators and the most straightforward example is $\exp({i\hat c_b^1})$.

This analysis leads to the conclusion that the mixture  between interior and exterior of the horizon by quantum effects is  a generic result in this quantization and that physical states entangle both regions.

\section{Conclusion}
\label{sec:concl}

We have argued that quantum mechanics applied to the whole spacetime generically introduces quantum correlations between  different classically disconnected regions (separated by horizons). This may be used as a first stage of an analysis of a quantum multiverse  scenario, in which  there may exist non-vanishing quantum correlations among individual otherwise uncorrelated universes. Ultimately, this would lead to the necessity of considering the whole multiverse in order to obtain a complete knowledge of our own universe.  

We have analyzed a Kantowski-Sachs minisuperspace model of a spacetime with a positive cosmological constant, whose classical solutions are Schwarzschild-de Sitter universes. We have carried out a canonical quantization of this model following (an extension of) Dirac's canonical quantization program for systems with first-class constraints. In this construction, the physical structure is  consistent and robust only if we consider the whole spacetime. We have proved that we can not restrict ourselves to the observed classical region when we consider the spacetime quantum mechanically, because there appear generically unavoidable quantum correlations between regions classically separated by horizons. This is explicitly shown by checking that unitarity is preserved only when the whole spacetime is taken into account. Indeed,  we have decomposed the physical Hilbert space of the system into two subspaces corresponding to states with support either between or beyond the horizons. These Hilbert subspaces are not stable under the action of unitary operators that describe a natural concept of evolution on physical states. Therefore, these states are better conceived as belonging to the tensor product of both subspaces, which are not separable but entangled.   

In contrast with many discussions carried out in quantum field theory on curved backgrounds (see e.g. \cite{ Unruh:1976db,Hawking:1974sw,Wald:1975kc,Israel:1976ur,Gibbons:1977mu,Einhorn:2005nw,MartinMartinez:2010ds,BirrelDavies}), a distinctive feature of our analysis is that our conclusions rest exclusively on the quantum behavior of the geometry. The entanglement between the regions in the interior and the exterior of the horizons has been shown to occur without introducing any field in the system: it is due solely to quantum properties of geometric observables on physical states of the Kantowski-Sachs model. Immediately, a new avenue is opened: Extending our investigations to the quantization of fields --for instance, a scalar one-- propagating on the quantum background studied here (this philosophy is similar to the strategy followed in the hybrid quantization scheme of Loop Quantum Cosmology. See \cite{MartinBenito:2008ej,FernandezMendez:2012vi,FMO13,CFMO14}). Then one could study perturbations of homogeneous (i.e., only $r$-dependent) scalar fields on this minisuperspace. In order to treat the background minisuperspace exactly, the ``zero mode'' of the scalar field (describing its homogeneous part) could be set to zero. Then we could expand the genuine perturbations of the field in a mode basis, for which one can consider e.g. a generalization of the analysis of Ref. \cite{Pereira:2012ma}.

Within this framework, one could analyze the differences between two ways of quantizing the model with the field. The first one would be quantizing the field separately in the sector of positive $c_0$ and negative $c_0$ at $b_0$, and the second one quantizing it  on the whole real line for $c_0$. This would allow us to look for quantum-field-theory effects and entropy mixing between both regions, but incorporating in the discussion the quantum nature of the geometry. In this manner, one would extend to the realm of quantum spacetime previous studies in the localization of quantum modes in a cavity, in which the tension between vacuum entanglement and having localized states clearly shows up \cite{Rodriguez-Vazquez:2014hka}.

\begin{acknowledgments}

A. A-S. is grateful to A. Ashtekar, M. Fern\'andez-M\'endez, and P.  Mart\'{\i}n-Moruno for conversations. The authors acknowledge financial support from the research grant MICINN/MINECO FIS2011-30145-C03-02 from Spain.
\end{acknowledgments}


\begin{thebibliography}{99.}

\bibitem{Jacobson:1991gr}
 T.~Jacobson,
 %``Black hole evaporation and ultrashort distances,''
 Phys. Rev. D {\bf 44}, 1731 (1991).

\bibitem{Jacobson:1993hn}
 T.~Jacobson,
 %``Black hole radiation in the presence of a short distance cutoff,''
 Phys. Rev. D  {\bf 48}, 728 (1993).
  
\bibitem{Unruh:1994je}
 W.~G.~Unruh,
 %``Sonic analog of black holes and the effects of high frequencies on black hole evaporation,''
 Phys. Rev. D {\bf 51}, 2827 (1995).

\bibitem{Unruh:1976db}
  W.~G.~Unruh,
  % `Notes on black hole evaporation,''
  Phys.\ Rev. D {\bf 14}, 870 (1976). 
    
\bibitem{Barbado:2011ai}
  L.C.~Barbado, C.~Barcel\'o, L.J.~Garay, and G.~Jannes,
  %``The Trans-Planckian problem as a guiding principle,''
   JHEP {\bf 1111}, 112 (2011).
       
\bibitem{Morris:1988tu}
  M.S.~Morris, K.S.~Thorne, and U.~Yurtsever,
  %``Wormholes, Time Machines, and the Weak Energy Condition,''
  Phys.\ Rev.\ Lett.\  {\bf 61}, 1446 (1988). 
 
\bibitem{Morris:1988cz}
  M.S.~Morris and K.S.~Thorne,
  %``Wormholes in space-time and their use for interstellar travel: A tool for teaching general relativity,''
  Am.\ J.\ Phys.\  {\bf 56}, 395 (1988). 
  
\bibitem{GonzalezDiaz:1996sr}
  P.F.~Gonz\'alez-D\'{\i}az,
  %``Ring holes and closed timelike curves,''
  Phys.\ Rev.\ D {\bf 54}, 6122 (1996).
  
\bibitem{Yurov:2006we}
  A.V.~Yurov, P.~Mart\'{\i}n Moruno, and P.F.~Gonz\'alez-D\'{\i}az,
  %``New Bigs in cosmology,''
  Nucl.\ Phys.\ B {\bf 759}, 320 (2006).
      
\bibitem{Gott:1997pm}
  J.R.~Gott III and L.-X.~Li,
  %``Can the universe create itself?,''
  Phys.\ Rev.\ D {\bf 58}, 023501 (1998).

\bibitem{GonzalezDiaz:1998pc}
  P.F.~Gonz\'alez-D\'{\i}az,
  %``Perdurance of multiply connected de Sitter space,''
  Phys.\ Rev.\ D {\bf 59}, 123513 (1999).

\bibitem{Hawking:1988ae}
  S.W.~Hawking,
  %``Wormholes in Space-Time,''
  Phys.\ Rev.\ D {\bf 37}, 904 (1988). 

\bibitem{Giddings:1988wv}
  S.B.~Giddings and A.~Strominger,
   %``Baby Universes, Third Quantization and the Cosmological Constant,''
Nucl.\ Phys.\ B {\bf 321}, 481 (1989). 
   
\bibitem{Hawking:1990in}
  S.W.~Hawking and D.N.~Page,
  %``The spectrum of wormholes,''
  Phys.\ Rev.\ D {\bf 42}, 2655 (1990). 
 
\bibitem{Garay:1993rr}
   L.J.~Garay,
   %``Hilbert space of wormholes,''
   Phys.\ Rev.\ D {\bf 48}, 1710 (1993).
 
\bibitem{MenaMarugan:1994qz}
   G.A.~Mena Marug\'an,
   %``Wormholes as basis for the Hilbert space in Lorentzian gravity,''
   Phys.\ Rev.\ D {\bf 50}, 3923 (1994).
  
\bibitem{GonzalezDiaz:1997xc}
   P.F.~Gonz\'alez-D\'{\i}az,
   %``Observable effects from space-time tunneling,''
   Phys.\ Rev.\ D {\bf 56}, 6293 (1997).
      
\bibitem{Shatskiy:2007xr}
  A.~Shatskiy,
  %``Passage of Photons Through Wormholes and the Influence of Rotation on the Amount of Phantom Matter around Them,''
  Astron.\ Rep.\  {\bf 51}, 81 (2007).
  
\bibitem{Shatskiy:2008ym}
  A.~Shatskiy,
  %``Image of another universe being observed through a wormhole throat,''
  Phys.\ Usp.\  {\bf 52}, 811 (2009).

\bibitem{GonzalezDiaz:2011ke}
  P.F.~Gonz\'alez-D\'{\i}az and A.~Alonso-Serrano,
  %``Observing other universe through ringholes and Klein-bottle holes,''
  Phys.\ Rev.\ D {\bf 84}, 023008 (2011).
  
 \bibitem{RoblesPerez:2010ut}
   S.~Robles-P\'erez and P.F.~Gonz\'alez-D\'{\i}az,
   %``Quantum state of the multiverse,''
   Phys.\ Rev.\ D {\bf 81}, 083529 (2010).

\bibitem{RoblesPerez:2011yj}
  S.~Robles-P\'erez, A.~Alonso-Serrano, and P.F.~Gonz\'alez-D\'{\i}az,
  %``Decoherence in an accelerated universe,''
  Phys.\ Rev.\ D {\bf 85}, 063511 (2012).
  
 \bibitem{RoblesPerez:2011zp}
   S.~Robles-P\'erez and P.F.~Gonz\'alez-D\'{\i}az,
   %``Quantum entanglement in the multiverse,''
   J.\ Exp.\ Theor.\ Phys.\  {\bf 118}, 34 (2014).

 \bibitem{AlonsoSerrano:2012wc}
   A.~Alonso-Serrano, C.~Bastos, O.~Bertolami, and S.~Robles-P\'erez,
   %``Interacting universes and the cosmological constant,''
   Phys.\ Lett.\ B {\bf 719}, 200 (2013).
 
\bibitem{Coleman:1988tj} 
  S.R.~Coleman,
  %``Why There Is Nothing Rather Than Something: A Theory of the Cosmological Constant,''
  Nucl.\ Phys.\ B {\bf 310}, 643 (1988).
  
\bibitem{Halliwell:1990tu}
  J.~J.~Halliwell and J.~Louko,
  %``Steepest Descent Contours in the Path Integral Approach to Quantum Cosmology. 3. A General Method With Applications to Anisotropic Minisuperspace Models,''
  Phys.\ Rev.\ D {\bf 42}, 3997 (1990).
 
\bibitem{Campbell:1990uu}
  L.M.~Campbell and L.J.~Garay,
  %``Quantum Wormholes In Kantowski-sachs Space-time,''
  Phys.\ Lett.\ B {\bf 254}, 49 (1991).
  
\bibitem{Garay:1991um}
 L.J.~Garay,
 %``Quantum state of wormholes and path integral,''
 Phys.\ Rev.\ D {\bf 44}, 1059 (1991).

\bibitem{MenaMarugan:1993rs}
  G.A.~Mena Marug\'an,
  %``Reality conditions in nonperturbative quantum cosmology,''
  Class.\ Quant.\ Grav.\  {\bf 11}, 589 (1994).
 
\bibitem{MenaMarugan:1994sj}
  G.A.~Mena Marug\'an,
  %``Is the exponential of the Chern-Simons action a normalizable physical state?,''
  Class.\ Quant.\ Grav.\  {\bf 12}, 435 (1995).

\bibitem{Kuchar:1994zk}
  K.V.~Kucha\v r,
  %``Geometrodynamics of Schwarzschild black holes,''
  Phys.\ Rev.\ D {\bf 50}, 3961 (1994). 

\bibitem{Gambini:2013ooa}
  R.~Gambini and J.~Pullin,
  %``Loop quantization of the Schwarzschild black hole,''
  Phys.\ Rev.\ Lett.\  {\bf 110}, 211301 (2013). 
  
\bibitem{yeshiva} P.A.M.~Dirac, {\it Lectures on Quantum Mechanics} (Yeshiva University, 
	New York, 1964).

\bibitem{Ashtekar:1994kv} 
  A.~Ashtekar and R.S.~Tate,
  %``An Algebraic extension of Dirac quantization: Examples,''
  J.\ Math.\ Phys.\  {\bf 35}, 6434 (1994).

\bibitem{Lake:2005bf}
  K.~Lake,
  %``A Maximally extended, explicit and regular covering of the Schwarzschild - de Sitter vacua in arbitrary dimension. I. General formulae,''
  Class.\ Quant.\ Grav.\  {\bf 23}, 5883 (2006). 
    
\bibitem{Ashtekar:2006uz}
  A.~Ashtekar, T.~Pawlowski, and P.~Singh,
  %``Quantum Nature of the Big Bang: An Analytical and Numerical Investigation. I.,''
  Phys.\ Rev.\ D {\bf 73},  124038 (2006).
 
\bibitem{Hawking:1974sw}
  S.~W.~Hawking,
  %``Particle Creation by Black Holes,''
  Commun.\ Math.\ Phys.\  {\bf 43}, 199 (1975)
   [Erratum-ibid.\  {\bf 46}, 206 (1976)].
   
\bibitem{Wald:1975kc}
  R.M.~Wald,
  %``On Particle Creation by Black Holes,''
  Commun.\ Math.\ Phys.\  {\bf 45}, 9 (1975).
  
\bibitem{Israel:1976ur}
  W.~Israel,
  %``Thermo field dynamics of black holes,''
  Phys.\ Lett.\ A {\bf 57}, 107 (1976).
   
\bibitem{Gibbons:1977mu}
  G.W.~Gibbons and S.W.~Hawking,
  %``Cosmological Event Horizons, Thermodynamics, and Particle Creation,''
  Phys.\ Rev.\ D {\bf 15}, 2738 (1977).
  
\bibitem{Einhorn:2005nw}
  M.B.~Einhorn and M.~Mahato,
  %``Beyond the horizon,''
  Phys.\ Rev.\ D {\bf 73}, 104035 (2006).
  
\bibitem{MartinMartinez:2010ds}
  E.~Mart\'{\i}n-Mart\'{\i}nez and J.~Le\'on,
  %``Quantum correlations through event horizons: Fermionic versus bosonic entanglement,''
  Phys.\ Rev.\ A {\bf 81}, 032320 (2010).
  
\bibitem{BirrelDavies} N.D. Birrell and P.C.W. Davies, {\it Quantum Fields in Curved Space} (Cambridge University Press, Cambridge, 1984).

\bibitem{MartinBenito:2008ej}
 M.~Mart\'{\i}n-Benito, L.J.~Garay, and G.A.~Mena Marug\'an,
  %``Hybrid Quantum Gowdy Cosmology: Combining Loop and Fock Quantizations,''
   Phys.\ Rev.\ D {\bf 78}, 083516 (2008).
   
\bibitem{FernandezMendez:2012vi}
  M. Fern\'andez-M\'endez, G.A. Mena Marug\'an, and J. Olmedo,
  %``Hybrid quantization of an inflationary universe,''
  Phys. Rev. D {\bf 86}, 024003 (2012).

\bibitem{FMO13}    
  M. Fern\'andez-M\'endez, G.A. Mena Marug\'an, and J. Olmedo,  
  % ``Hybrid quantization of an inflationary model: The flat case,''
  Phys. Rev. D {\bf 88}, 044013 (2013).
  
\bibitem{CFMO14}  
  L. Castell\'o Gomar, M. Fern\'andez-M\'endez, G.A. Mena Marug\'an, and J. Olmedo, 
  %``Cosmological perturbations in hybrid loop quantum cosmology: Mukhanov-Sasaki variables,''
  Phys. Rev. D {\bf 90}, 064015 (2014).
   
\bibitem{Pereira:2012ma}
  T.S.~Pereira, S.~Carneiro, and G.A.~Mena~Marug\'an,
  %``Inflationary Perturbations in Anisotropic, Shear-Free Universes,''
  JCAP {\bf 1205} (2012) 040.

\bibitem{Rodriguez-Vazquez:2014hka}
  M.~Rodr\'{\i}guez-V\'azquez, M.~del Rey, H.~Westman, and J.~Le\'on,
  %``Local quanta, unitary inequivalence, and vacuum entanglement,''
   Annals  Phys.   {\bf 351}, 112 (2014). 
   
 
  \end{thebibliography}
  \end{document}